\documentclass[reprint,prb,amsmath,amssymb,letterpaper]{revtex4-1}
\usepackage{graphicx}
\usepackage{bm}
\usepackage{amssymb} 
\usepackage{amsmath}
\usepackage{dcolumn}

\begin{document}

\title{Fluorescence quenching near small metal nanoparticles}

\author{V. N. Pustovit$^{1}$ and T. V. Shahbazyan$^{2}$}

\affiliation{$^{1}$Centre de Recherche Paul Pascal, CNRS UPR 8641, 115 Avenue Albert Schweitzer, 33600 Pessac, France\\
$^{2}$Department of Physics, Jackson State University, Jackson, MS 39217, USA}


\begin{abstract}
We develop a microscopic model for fluorescence of a molecule (or semiconductor quantum dot) near a small metal nanoparticle.  When a molecule is situated close to metal surface, its fluorescence is quenched due to energy transfer to the metal. We perform quantum-mechanical calculations of energy transfer rates for nanometer-sized Au nanoparticles and find that non-local and quantum-size effects significantly enhance dissipation in metal as compared to those predicted by semiclassical electromagnetic models. However, the dependence of transfer rates on molecule's distance to metal nanoparticle surface, $d$, is significantly weaker than the $d^{-4}$ behavior for flat metal surface with a sharp boundary predicted by previous calculations within random phase approximation.

\end{abstract}


\maketitle


\section{Introduction}

During past decade, fluorescence of molecules placed near a metal nanostructure supporting surface plasmon (SP) has attracted renewed interest due to biosensing applications.\cite{dubertret-naturebiotech01,nie-jacs02,vanduyne-nm06,ray-plasm07,mayer-cr11} While early studies mainly focused on fluorescence near rough metal films, \cite{brus-jcp80,ritchie-prb81,weitz-ol82,alivisatos-jcp85} more recent advances in near-field optics and in chemical control of molecule-nanostructure complexes spurred a number of fluorescence measurements from dye molecules and semiconductor quantum dots (QDs) linked to metal nanoparticles (NP).\cite{feldmann-prl02,lakowicz-jf02,artemyev-nl02,gueroui-prl04,feldmann-nl05,klimov-jacs06,mertens-nl06,pompa-naturenanotech06,novotny-prl06,sandoghdar-prl06,novotny-oe07,sandoghdar-nl07,lakowicz-jacs07,chen-nl07,lakowicz-nl07,halas-nl07,feldmann-nl08,halas-acsnano09,ming-nl09,kinkhabwala-naturephot09,viste-acsnano10,lakowicz-jacs10,munechika-nl10,ming-nl11,ratchford-nl11,raino-acsnano11} Fluorescence quantum efficiency of such conjugate systems is determined by competition between strong local field enhancement of molecule's dipole moment near SP resonance\cite{moskovits-rmp85} and, hence, of radiative decay rate $\Gamma_{r}$, and nonradiative energy transfer from excited dipole to optically-inactive electronic excitations in the metal characterized by nonradiative decay rate $\Gamma_{nr}$.\cite{silbey-acp78,metiu-pss84} The interplay between these two mechanisms depends on separation, $d$, of the emitter from the metal surface.\cite{nitzan-jcp81,ruppin-jcp82} As a result, fluorescence is enhanced at some optimal $d$ and is quenched close to NP due to the suppression of quantum efficiency, $Q=\Gamma_{r}/\left (\Gamma_{r}+\Gamma_{nr}\right )$, by non-radiative processes. Both enhancement and quenching were widely observed in fluorescence experiments on Au and Ag nanoparticles.\cite{feldmann-prl02,lakowicz-jf02,artemyev-nl02,gueroui-prl04,feldmann-nl05,klimov-jacs06,mertens-nl06,pompa-naturenanotech06,novotny-prl06,sandoghdar-prl06,novotny-oe07,sandoghdar-nl07,lakowicz-jacs07,chen-nl07,lakowicz-nl07,halas-nl07,feldmann-nl08,halas-acsnano09,ming-nl09,kinkhabwala-naturephot09,viste-acsnano10,lakowicz-jacs10,munechika-nl10,ming-nl11,ratchford-nl11,raino-acsnano11}

When a dipole with excitation energy $\omega$ is situated near a spherical NP of radius $R$ its energy is transferred to via Coulomb interactions to electronic excitations in NP (single-particle or collective) with angular momentum $l$. Within classical approach,\cite{nitzan-jcp81,ruppin-jcp82} these are collective modes described by resonances of NP $l$-pole  dynamic polarizabilities $\alpha_{l}(\omega)$ expressed in terms of \textit{local} dielectric function $\epsilon(\omega,{\bf r})$, i.e.,  bulk metal dielectric function $\epsilon(\omega)=\epsilon'(\omega)+i\epsilon''(\omega)$ for $r<R$ and dielectric constant of outside medium $\epsilon_{m}$ for $r>R$. The energy dissipation in NP is then determined by $\epsilon''(\omega)$  which comes mainly from electron-phonon interactions and, for high $\omega$, from optical transitions between d-band and sp-band. The distance dependence of energy transfer rate $\Gamma_{nr}$ stems from electromagnetic interaction of molecular dipole with its image that, at small $d$ (i.e., $d/R\ll 1$), yields\cite{nitzan-jcp81,ruppin-jcp82} $\Gamma_{nr}\propto d^{-3}$, similar to flat metal surface.\cite{silbey-acp78}  Recent fluorescence measurements from molecules near (but not very close to) relatively large (tens of nm) NPs \cite{novotny-prl06,sandoghdar-prl06,novotny-oe07,sandoghdar-nl07} indicated a good agreement with classical electromagnetic models.\cite{nitzan-jcp81,ruppin-jcp82,carminati-oc06,mertens-prb07}

With decreasing distance between dipole and metal surface, the energy transfer rate gets significantly modified due to spatial dispersion of $\epsilon$. For $d\lesssim v_{F}/\omega$, where $v_{F}$ is Fermi velocity, the high Fourier harmonics of dipole's Coulomb potential are no longer screened by conduction electrons in the metal so that the phase space restriction on excitation of electron-hole pairs over Fermi level is lifted.\cite{silbey-acp78} For small $d$, electron-hole generation becomes the dominant energy transfer mechanism that leads to a sharper distance dependence of energy transfer rate, $\Gamma_{nr}\propto d^{-4}$; the precise coefficient calculated within random phase approximation (RPA) differs in the literature.\cite{persson-prb82,stockman-prb04} 

While fluorescence quenching from a molecule near a relatively large NP is essentially similar to that near flat metal surface,\cite{kall-prl04,kall-prb05} the energy transfer to \textit{small} NPs is significantly modified due to quantum-size effects.  Fluorescence near nanometer-sized Au NPs with diameter as small as 1.4 nm has attracted significant interest due to their large surface to volume ratio and, hence, lower Ohmic losses, which is important for biosensing applications.\cite{dubertret-naturebiotech01,nie-jacs02,raino-acsnano11} For such NPs, a spillover of a substantial fraction of electron density beyond NP classical radius can significantly affect spatial dispersion of metal dielectric function \cite{ekardt-prb85,persson-prb85,koch-prb96}  and hence modify the dipole Coulomb potential near NP surface. The presence of strongly localized d-band with abrupt nearly classical density profile gives rise to a surface layer where Coulomb potential is not screened by interband transitions.\cite{liebsch-prb93,liebsch-prb95,kresin-prb95,lerme-prl98,voisin-prl00} In small NPs, the size-dependent Landau damping of plasmons excited by the molecular dipole can increase the rate of energy dissipation.\cite{kreibig-book} Incorporation of all the above effects in a single model requires a microscopic approach of energy transfer rates near small NPs.

In this paper, we present such a microscopic model, based on time-dependent local density approximation (TDLDA), for calculation of radiative and nonradiative decay rates of a molecular dipole near a small noble metal NP. Here, the molecule is represented by a point-like dipole while we focus on non-local and quantum-size effects in dielectric function of nanometer-sized metal NP. To this end, we adopt quantum three-region model we developed recently\cite{pustovit-prb06,pustovit-epj09} that accounts for different density profiles of sp-band and d-band electrons \cite{liebsch-prb93,liebsch-prb95,kresin-prb95,lerme-prl98,voisin-prl00} in small noble metal NP embedded in a dielectric medium. We find that calculated radiative decay rates, $\Gamma_{r}$, are not significantly affected by non-local effects and are comparable to those calculated using semiclassical approaches\cite{nitzan-jcp81,ruppin-jcp82} with quantum corrections. At the same time, the energy transfer rate $\Gamma_{nr}$ is significantly enhanced due to the generation of electron-hole pairs in metal at small molecule-NP separations. We find, however, that quantum-size and non-local effects strongly affect the distance dependence of $\Gamma_{nr}$ close to NP surface and that $d^{-4}$ behavior no longer holds; namely, the energy transfer rate is significantly slower than that predicted by RPA calculations.\cite{persson-prb82,stockman-prb04}

The paper is organized as follows. In Section \ref{sec:dipole} we set up the expressions for decay rates to be evaluated and derive the classical limit. In Section \ref{sec:tdlda} we outline the quantum three-region model used for evaluation of NP polarizabilities. In Section \ref{sec:numeric} our numerical results are presented and discussed, and Section \ref{sec:conc} concludes the paper.

\section{Radiation of a dipole near spherical metal nanoparticle}
\label{sec:dipole}

Consider a molecule with a dipole moment $\bm{\mu}$ located at position $\bf{r}$ measured  from the center of a metal NP of radius $R$ placed in a medium with dielectric constant $\epsilon_{m}$.  Its decay rate  is given by a general expression \cite{novotny-book}
\begin{equation}
\label{green}
\Gamma=4\pi k^{2}\, {\rm Im}\, \bm{\mu}\cdot {\bf G(r,r)}\cdot\bm{\mu},
\end{equation}
where $G_{\mu\nu}({\bf r},{\bf r}')$ is the electric field Green dyadic in the presence of NP and $k$ is the wave vector of light. The full decay rate Eq. (\ref{green} includes contributions coming from two distinct processes, $\Gamma=\Gamma_{r}+\Gamma_{nr}$: emission of a photon by NP-dipole system with the rate $\Gamma_{r}$, and energy transfer to electronic excitations in the NP  with the rate $\Gamma_{nr}$. If the system size is  sufficiently small, $kr\ll 1$, then $\Gamma_{nr}$ can be found in a standard way from the longwave  limit of the Green dyadic,\cite{novotny-book} 
\begin{equation}
G_{\mu\nu}({\bf r},{\bf r}')=-\frac{-1}{4\pi k^{2}}\,\nabla_{\mu}\nabla'_{\nu} U({\bf r},{\bf r'})
\end{equation}
where  
%
\begin{equation}
\label{coulomb}
U({\bf r}, {\bf r'})
   =\sum_{lm} \frac{4\pi}{2l+1}
\left[\frac{r_{<}^{l}}{r_{>}^{l+1}}
-\frac{\alpha_l}{\left(rr'\right)^{l+1}} \right]
Y_{lm}(\hat{\bf r})Y_{lm}^{\ast}(\hat{\bf r}'),
\end{equation}
is expansion of Coulomb potential in the presence of NP over spherical harmonics $Y_{lm}(\hat{\bf r})$, and $r_{>}$ and $r_{<}$ are the larger and smaller values of $r$ and $r'$, respectively. The second term in Eq.~(\ref{coulomb}) contains NP $l$-pole dynamic polarizability, $\alpha_{l}(\omega)$, given by
\begin{equation}
\label{polar}
\alpha_{l}= \frac{4\pi}{2l+1}\int dr r^{l+2}\delta n^{(l)}(r),
\end{equation}
where $\delta n^{(l)}(r)$ is the induced density due to potential $r^{l}$ (hereafter we suppress $\omega$-dependence to alleviate expressions). From Eqs.~(\ref{green}) and (\ref{coulomb}),
$\Gamma_{nr}$  can be calculated for arbitrary dipole orientation with respect to the NP surface; for perpendicular and parallel orientations, it takes the form \cite{nitzan-jcp81,ruppin-jcp82}
\begin{align}
\label{nonrad}
\Gamma_{nr}^{\perp}=\frac{3\Gamma_{r}^{0}}{2k^{3}}\sum_{l} \frac {(l+1)^{2} \alpha''_l}{r^{2l+4}}, 
~
\Gamma_{nr}^{\parallel}=\frac{3\Gamma_{r}^{0}}{2k^{3}}\sum_{l} \frac {l(l+1) \alpha''_l}{2r^{2l+4}},
\end{align}
where $\Gamma_{r}^{0}=\frac{2}{3}\mu^{2}k^{3}$ is radiative decay rate of an isolated dipole. The radiative decay rates in the presence of NP are similarly obtained using Green dyadic's far field asymptotics as \cite{nitzan-jcp81,ruppin-jcp82} 
\begin{align}
\label{rad}
\Gamma_{r}^{\perp}=\Gamma_{r}^{0}\left |1+ 2\frac { \alpha_{1}}{r^{3}}\right|^{2},
~~
\Gamma_{r}^{\parallel}=\Gamma_{r}^{0}\left |1 - \frac { \alpha_{1}}{r^{3}}\right|^{2},
\end{align}
and they depend on NP dipole polarizability, $\alpha_{1}(\omega)$, that peaks at surface plasmon frequency, $\omega_{sp}$. Note that the expression (\ref{polar}) for NP polarizability that determines decay rates (\ref{nonrad}) and (\ref{rad})  is valid as long as there is no direct overlap between molecule orbitals and electron states in the metal.\cite{pustovit-prb06} In classical approximation of a sharp NP boundary at $R$, the induced density $\delta n_{l}(r)$ peaks at $r=R$, yielding standard expression for NP polarizability, 
\begin{equation}
\label{polar-classical}
\alpha_{l}=R^{2l+1}\frac{\epsilon - \epsilon_{m}}{\epsilon + \epsilon_{m}(1+l^{-1})},
\end{equation}
where $\epsilon(\omega)$ is the metal dielectric function. As the dipole distance to NP surface, $d=r-R$, decreases, $\Gamma_{r}$ and $\Gamma_{nr}$ exhibit significantly different behavior.  The distance dependence of $\Gamma_{r}$ is determined largely by the interference between direct and NP-induced terms, but it remains finite for $d/R\ll 1$. In contrast, $\Gamma_{nr}$ sharply increases for small $d$ due to dominant contribution of high-$l$ terms in Eq.~(\ref{nonrad}). Indeed, using Drude form for bulk metal dielectric function, the large $l$ approximation for NP polarizability (\ref{polar-classical}) is $\alpha''_{l}\sim \omega \gamma_{b}/\omega_{b}^{2} $, where $\omega_{b}$ and $\gamma_{b}$ are, respectively, bulk plasmon energy and damping rate. For $d/R\ll 1$, the estimate of $\Gamma_{nr}$ then takes the form
\begin{equation}
\Gamma_{nr}\sim \frac{\mu^{2}}{R^{3}}\frac{\omega\gamma_{b}}{\omega_{b}^{2}}\int \frac{dl l^{2}}{ (1+d/R)^{2l+4}}\sim \frac{\mu^{2}}{d^{3}}\frac{\omega\gamma_{b}}{\omega_{b}^{2}}, 
\end{equation}
where $l$ is restricted by $l_{max}\sim k_FR \gg 1$, $k_F$ being the electron Fermi momentum. In the local approximation when the spatial dispersion of metal dielectric function $\epsilon(\omega)$ is neglected, $\gamma_{b}$ is determined by electron bulk scattering length, $l_{sc}$ due to mainly electron-phonon and impurity scattering, $\gamma_{b}\sim v_{F}/l_{sc}$, resulting in $\Gamma_{nr}\propto d^{-3}$ distance dependence. The  RPA calculations non-local corrections at small $d$ due to electron-hole pairs generation\cite{persson-prb82,stockman-prb04} change this behavior to $\Gamma_{nr}\propto d^{-4}$. In the rest of the paper, we present microscopic calculation of the decay rates and fluorescence quantum efficiency.

\section{Quantum three-region model for noble metal nanoparticle polarizabilities}
\label{sec:tdlda}

For calculation of NP polarizabilities Eq.~(\ref{polar}), we adopt quantum three-region model developed by us previously \cite{pustovit-prb06,pustovit-epj09} that combines a quantum-mechanical description for {\em sp}-band electrons within TDLDA and phenomenological treatment of {\em d}-electrons with dielectric function $\epsilon_{d}(\omega)$ in the region confined by $R_d<R$ and of outside medium with dielectric constant $\epsilon_{m}$. This model represents an extension of two-region model (i.e., without outside dielectric) that is known to describe reasonably well screening effects for symmetric (closed shell) spherical clusters with electron numbers ranging from several dozens to several thousands.\cite{liebsch-prb93,liebsch-prb95,kresin-prb95,lerme-prl98,voisin-prl00} In the longwave approximation, local fields at point ${\bf r}$ outside NP are determined by quasistatic potential $\phi({\bf r},\omega)=\phi_{0}({\bf r})+\delta\phi({\bf r},\omega)$, where 
\begin{eqnarray}
\label{poisson}
\delta\phi (\omega,{\bf r})
=e^2 \int d^3r' \,
\frac{\delta n(\omega,{\bf r}')}{|{\bf r}-{\bf r}'|},
\end{eqnarray}
is induced potential. The induced density is decomposed as $\delta n({\bf r})= \delta n_s({\bf r}) +\delta n_d({\bf r}) +\delta n_m({\bf r})$, where $\delta n_s({\bf r})$, $\delta n_d({\bf r})$, and $\delta n_m({\bf r})$ are, respectively, contributions from {\em sp}-electrons, {\em d}-electrons, and surrounding medium. The {\em sp}-electrons contribution is determined from TDLDA equation \cite{ekardt-prb85}
\begin{eqnarray}
\label{tdlda}
\delta n_s({\bf r}) = \int d^3 r' P_s ({\bf r}, {\bf r}') \Bigl[\phi({\bf r}')
+ V'_x[n(r')]\delta n_s ({\bf r}')\Bigr],
\end{eqnarray}
where $P_s ({\bf r}, {\bf r}')$ is polarization operator for noninteracting  {\em sp}-electrons, $V'_x[n(r')]$ is the (functional) derivative of  exchange-correlation potential,\cite{lundqvist-prb77} $n(r)$ being the ground-state electron density; $P_s ({\bf r}, {\bf r}')$ and $n(r)$ are calculated in a standard way using Kohn-Sham equations.\cite{koch-prb96} The system is closed by expressing  $\phi({\bf r})$ in Eq.~(\ref{tdlda}) via $\delta n_s({\bf r})$. This is accomplished by relating $\delta n_d({\bf r})$ and $\delta n_m({\bf r})$ in Eq.~(\ref{poisson}) back to $\phi({\bf r})$ as \cite{pustovit-prb06,pustovit-epj09}
\begin{eqnarray}
\label{back}
e^2 \delta n_d({\bf r})= \nabla
\bigl[\chi_d(r)\nabla \phi({\bf r})\bigr],
\nonumber\\
e^2 \delta n_m({\bf r})= 
\nabla \bigl[\chi_m(r)\nabla \phi({\bf r})\bigr],
 \end{eqnarray}
where
$\chi_d(r,\omega)=\bigl[(\epsilon_d(\omega)-1)/4\pi\bigr] \,\theta(R_d-r)$ is interband susceptibility with step function $\theta(x)$ enforcing the boundary conditions and, correspondingly, $\chi_m(r)=\bigl[(\epsilon_m-1)/4\pi\bigr]\,\theta(r-R)$ is the susceptibility of surrounding medium. Using spherical harmonics expansions $\phi({\bf r})=\sum_l \phi^{(l)}(r) Y_{lm}(\hat{\bf r})$ and $\delta n({\bf r})=\sum_l \delta n^{(l)}(r) Y_{lm}(\hat{\bf r})$ as well as the continuity of $\phi({\bf r})$ at $r=R$ and $r=R_{d}$, we obtain decomposition $\phi^{(l)}(r)=w_{0}^{(l)}(r)+\delta w_{0}^{(l)}(r) +\delta w_{s}^{(l)}(r)$, where $w_{0}^{(l)}=r^{l}/\epsilon(r)$ is potential of external field, $\delta w_{0}^{(l)}(r)$ is contribution from d-band and outside medium, and $\delta w_{s}^{(l)}(r)$ is sp-band contribution. The latter is related to the sp-band induced density as\cite{pustovit-prb06,pustovit-epj09}
\begin{equation}
\label{dws}
\delta w_{s}^{(l)}(r) = \int_0^{\infty}dr' r'^2 A^{(l)}(r,r') \delta n_{s}^{(l)}(r'),
\end{equation}
where the kernel $A^{(l)}(r,r')$ is given by
\begin{align}
\label{A}
A^{(l)}(r,r')=
 \frac{4\pi}{(2l+1)\bar{\epsilon}(r)} \left[ B_{0}^{(l)}(r,r') + B_{s}^{(l)}(r,r')\right]
 \end{align}
with $\bar{\epsilon}(r)= \epsilon_d$, 1, and $\epsilon _m$ for $r$ in regions $(0,R)$, $(R_d,R)$, and $(R,\infty)$. The functions $B_{0}^{(l)}(r,r')$  and $B_{s}^{(l)}(r,r')$ are, respectively, the $l$-th harmonics of direct and image Coulomb potential:
\begin{align}
\label{B0}
B_{0}^{(l)}(r,r') =\frac{r'^l}{r^{l+1}}\,\theta(r-r') +\frac{r^l}{r'^{l+1}}\,\theta(r'-r),
\end{align}
and
\begin{align}
\label{Bs}
&B_{s}^{(l)}(r,r')=
\beta_l(r/R)\frac{\lambda_m}{\eta}
\Bigl[B_{0}^{(l)}(R,r')
\nonumber\\
&-la^{l+1}\lambda_d B_{0}^{(l)}(R_d,r') \Bigr]
-\beta_l(r/R_d)\frac{\lambda_d}{\eta}\Bigl[B_{0}^{(l)}(R_d,r')
\nonumber\\
& \qquad \qquad \qquad \qquad \qquad
-a^{l}(l+1)
\lambda_m B_{0}^{(l)}(R,r')\Bigr],
\end{align}
where $\beta_l(r/R)=\frac{2l+1}{4\pi}R^2\partial_R B_{0}^{(l)}(R,r)$ is the derivative of Coulomb potential at the boundary, 
\begin{align}
\label{beta}
\beta_l(x) = lx^{-l-1}\, \theta (x-1) -(l+1)x^l \theta(1-x),
\end{align}
and coefficients $\lambda_d$, $\lambda_m$, $\eta$, and $a$ are given by
\begin{align}
\label{lambda}
\lambda_d=\frac{\epsilon_d-1}{l\epsilon_d+l+1}, ~~~
\lambda_m=\frac{\epsilon_m-1}{(l+1)\epsilon_m+1}, 
\nonumber\\
\eta=1-l(l+1)a^{2l+1}\lambda_d\lambda_m, ~~~
a=R_d/R.
\end{align}
 The combined d-band and outside medium contribution has the form
 \begin{align}
 \label{dw0}
\delta w_{0}^{(l)}(r) = \frac{R^{l}}{\bar{\epsilon}(r)\eta} 
\Bigl[\beta_l(r/R) \lambda_m(1-la^{2l+1}\lambda_d)
\nonumber\\
 -\beta_{l}(r/R_d) \lambda_{m} a^l\left [1-(l+1)\lambda_{m}\right ]
\Bigr].
\end{align}
We then obtain a closed equation for $\delta n_{s}^{(l)}$ as,
\begin{eqnarray}
\label{tdlda2}
\delta n_s^{(l)}(r) =\int d r' r'^2 P_s^{(l)} (r,r') 
\Biggl[w_{0}^{(l)}(r')+\delta w_{0}^{(l)}(r')
\qquad
\\
+
\int d r'' r''^2  
A^{(l)}(r',r'')\delta n_s^{(l)} (r'')+ V'_x(r')\delta n_s^{(l)} (r')\Biggr],
\nonumber
\end{eqnarray}
where the polarization operator $P_s^{(l)} (r,r')$ and the exchange potential\cite{lundqvist-prb77} $V'_x(r')$ are calculated from Kohn-Sham eigenfunctions.\cite{koch-prb96} Finally, NP polarizability (\ref{polar}) can be derived from the induced potential (\ref{poisson}) outside NP using the relation $\delta \phi^{(l)}=e^{2}\alpha_{l}/\epsilon_{m}r^{l+1}$. Using Eqs. (\ref{dws})  and (\ref{dw0}), we obtain $\alpha_{l}=\alpha_{0}^{(l)}+\alpha_{s}^{(l)}$, where
\begin{equation}
\label{alpha-0}
\alpha_{0}^{(l)}(\omega)
=l R^{2l+1}\left[  a^{2l+1} \lambda_d(1-\lambda_m) - \lambda_m 
\right]/\eta,
\end{equation}
and 
\begin{align}
\label{alpha-s}
\alpha_s^{(l)}(\omega)
=
\frac{4\pi }{(2l+1)}
\int_0^{\infty}dr'r'^{2}\varphi_{l}(r')\delta n_s^{(l)}(r')
\end{align}
with
\begin{align}
\label{varphi}
\varphi_{l}(r)
= r^{l}-
l R_d^{l+1} \frac{\lambda_d}{\eta} (1+l\lambda_{m})B_{0}^{(l)}(R_d,r)
\qquad
\nonumber\\
 + l R^{l+1}\frac{\lambda_m}{\eta} \left [1+(l+1)a^{2l+1}\lambda_{d}\right ]
 B_{0}^{(l)}(R,r).
\end{align}
Equations (\ref{A})--(\ref{varphi}) determine self-consistently quantum  polarizabilities $\alpha_{l}(\omega)$ of noble metal NP in dielectric medium which, in turn, determine the decay rates (\ref{nonrad})  and (\ref{rad}).  In the next section, we present the results of our numerical calculations of $\Gamma_{r}$ and $\Gamma_{nr}$ and of fluorescence quantum efficiency for a molecule near a small NP. 

  \begin{figure}[tb]
 \begin{center}
  \includegraphics[width=0.95\columnwidth]{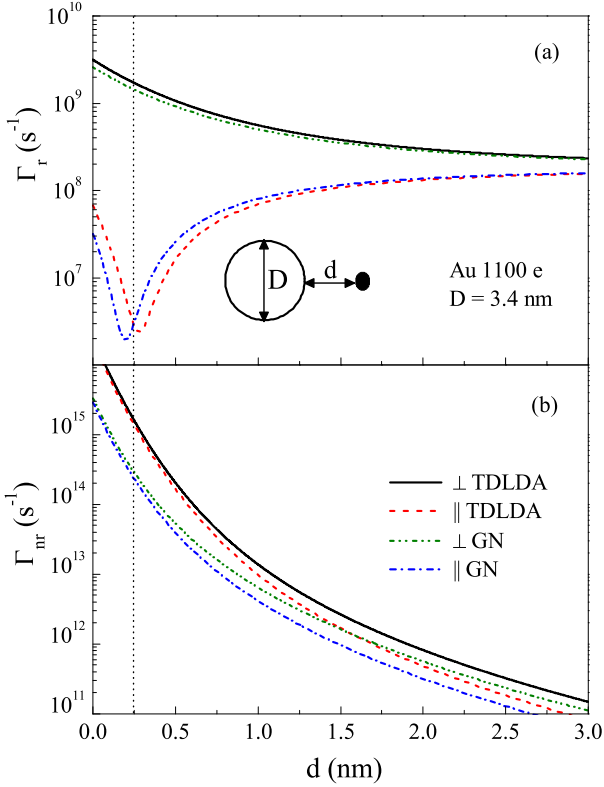}
 \end{center}
  \caption{\label{fig:1100rate} (Color online) Calculated radiative (a) and nonradiative (b) decay 
rates of a molecule near  $D=3.4$ nm diameter Au NP for normal and parallel dipole orientation are compared to semiclassical (Gersten-Nitzan) decay rates.}
  \end{figure}

  \begin{figure}[tb]
 \begin{center}
  \includegraphics[width=0.95\columnwidth]{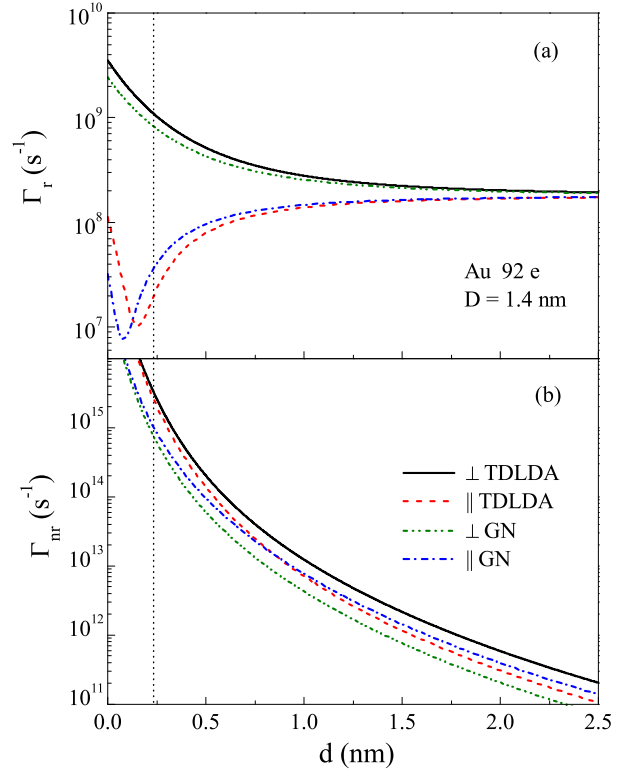}
  \end{center}
  \caption{\label{fig:92rate} (Color online) 
Same as in Fig.~\ref{fig:1100rate}, but for $D=1.4$ nm diameter Au NP.}
  \end{figure}

\section{Numerical results and discussion}
\label{sec:numeric}

Calculations of decay rates (\ref{nonrad})  and (\ref{rad}) were carried out for a point dipole situated near Au NP in a medium with dielectric constant $\epsilon_{m}=1.77$. We chose dipole's emission frequency to coincide with that of Cy5 fluorophore, 1.83 eV, which lies at low energy side of Au NP SP resonance (1.98 eV), so that quantum size and non-local effects we are interested in are not being obscured by interband absorption.  To elucidate the role of quantum-size effects, the decay rates $\Gamma_{r}$ and $\Gamma_{nr}$ were calculated for \textit{two} spherical Au clusters with electron numbers $N=92$ and $N=1100$ corresponding to  NP diameters $D\approx 1.4$ nm and $D\approx 3.4$ nm, respectively, and to highlight the role of non-local and screening effects the results of microscopic calculations are compared to those obtained using semiclassical polarizabilities, Eq.~(\ref{polar-classical}) with the Landau damping of plasmons $\gamma_{s}=v_{F}/R$ included in \textit{both} quantum and classical NP polarizabilities.  Both normal and parallel dipole orientations were considered and multipolar polarizabilities $\alpha_{l}(\omega)$ with angular momenta up to $l=30$ were included. The intrinsic decay rates, $\Gamma_{r}^{0}=1.8\times 10^{8}$ s$^{-1}$ and  $\Gamma_{nr}^{0}=1.08\times 10^{9}$ s$^{-1}$ of Cy5 dye were included in the full decay rates, and the bulk interband dielectric function\cite{palik-book} $\epsilon_{d}(\omega)$ was used for determination of three-region model parameters.\cite{pustovit-prb06,pustovit-epj09}  


In Figs. \ref{fig:1100rate}  and \ref{fig:92rate}, we show calculated radiative decay rate $\Gamma_{r}$ and energy transfer rate $\Gamma_{nr}$ for $D=3.4$ nm and $D=1.4$ nm Au NPs, respectively, together with the corresponding semiclassical rates based on Gersten-Nitzan (GN) model\cite{nitzan-jcp81} defined by NP polarizabilities Eq.~(\ref{polar-classical}) with size-dependent Landau damping correction included. For radiative decay rate $\Gamma_{r}$, the non-local effects are not significant except in narrow surface region (denoted by dashed line) were NP radius is not well defined due to spillover effects. This result is expected because only $l=1$ excitations contribute to $\Gamma_{r}$ through NP dipole polarizability $\alpha_{1}(\omega)$ which does not account, due to the phase space restrictions, for direct excitation of electron-hole pairs. A slightly larger quantum $\Gamma_{r}$ for normal dipole orientation is due to the difference in d-band and sp-band electron density profiles in the NP surface layer; the spillover of conduction electrons (as opposed to strongly localized d-band electrons) leads to reduction interband screening and hence larger local fields in the vicinity of NP boundary.\cite{liebsch-prb93,liebsch-prb95,kresin-prb95,lerme-prl98,voisin-prl00} This effect is stronger for small NP due to larger volume fraction of underscreened region.\cite{pustovit-prb06,pustovit-josa06} For parallel dipole orientation, the overall $d$-dependence of $\Gamma_{r}$ is dominated by destructive interference between direct and scattered waves.\cite{feldmann-nl05} Note that for small NPs, the minimum is reached in the spillover region, but moves towards outer region with increasing NP size. 

  \begin{figure}[tb]
 \centering
  \includegraphics[width=0.95\columnwidth]{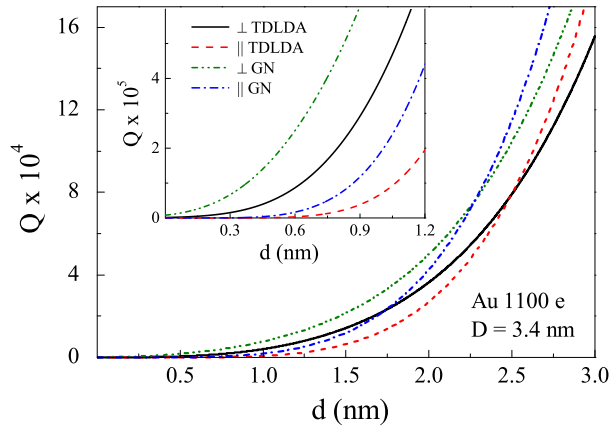}
  \caption{\label{fig:1100q} (Color online) Calculated fluorescence quantum efficiencies
near  $D=3.4$ nm diameter Au NP for normal and parallel dipole orientation are compared to semiclassical (Gersten-Nitzan) quantum efficiencies. Inset: blow-up for small molecule-NP distances.}
  \end{figure}

  \begin{figure}[tb]
  \centering
  \includegraphics[width=0.95\columnwidth]{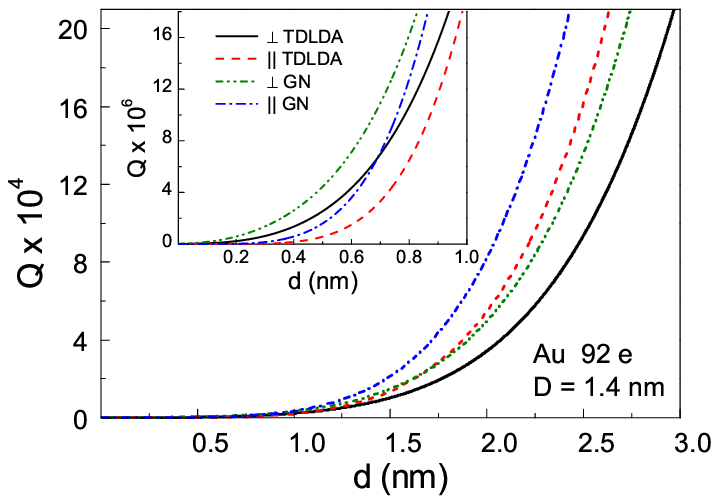}
  \caption{\label{fig:92q} (Color online) 
Same as in Fig.~\ref{fig:1100q}, but for $D=1.4$ nm diameter Au NP. }
  \end{figure}

  \begin{figure}[tb]
 \centering
  \includegraphics[width=0.95\columnwidth]{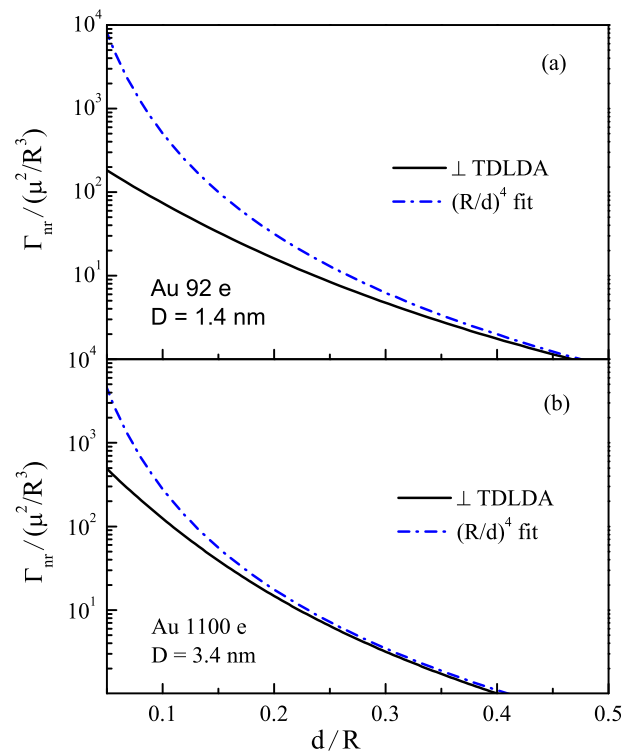}
  \caption{\label{fig:power} (Color online) Normalized nonradiative decay rates for $D=1.4$ nm (a) and $D=3.4$ nm diameter Au NP are shown together with $(d/R)^{4}$ fit.}
  \end{figure}

In contrast, the calculated quantum energy transfer rate $\Gamma_{nr}$ is significantly larger than its semiclassical counterpart for either dipole orientation. Close to NP surface, $\Gamma_{nr}$ for the larger $D=3.4$ nm NP is enhanced  by factor $\sim 10$ due to direct excitation of Fermi see electron-hole pairs by dipole's Coulomb potential\cite{persson-prb82,stockman-prb04} (see Fig.~\ref{fig:1100rate}). Note that the role of non-local effects in $\Gamma_{nr}$ considerably exceeds that of SP Landau damping which is included in both calculations. However, Landau damping is more important in small  $D=1.4$ nm NP, as it leads to smaller difference between quantum and semiclassical $\Gamma_{nr}$ while increasing the overall magnitude of $\Gamma_{nr}$ (see Fig.~\ref{fig:92rate}). This results in the overall larger fluorescence quantum efficiency for larger NPs, as shown in Figs.~\ref{fig:1100q} and \ref{fig:92q}. Although for relatively large NP-dipole distances ($d\gtrsim R$) quantum efficiencies $Q$ of 3.4 nm and 1.4. nm NPs are comparable, being determined mainly by intrinsic molecular decay rates, for small distances ($d/R\ll 1$) the larger NP $Q$ is significantly higher (see insets in  Figs.~\ref{fig:1100q} and \ref{fig:92q}). At the same time, the difference between quantum and semiclassical calculations of $Q$ is also larger for $D=3.4$ nm NP, indicating a larger role of non-local and screening effects. The latter point is illustrated in Fig.~(\ref{fig:power}) that compares the distance dependence of quantum $\Gamma_{nr}$ with $d^{-4}$ behavior near a flat surface predicted by RPA calculations.\cite{persson-prb82,stockman-prb04} In fact, the raise of $\Gamma_{nr}$ near NP surface is significantly slower for either NP size, but for the smaller $D=1.4$ nm NP the deviation from $d^{-4}$ behavior is considerably larger. 

As a final remark, the theory presented here is valid if there is no significant overlap between molecular orbitals and electron wave functions in the metal. This condition no longer holds when molecule-NP distance becomes several \AA, i.e., in the spillover region (vertical dashed lines in Figs. \ref{fig:1100rate} and \ref{fig:92rate}). In this region, fluorescence quenching is dominated by electron tunneling rather than energy transfer. We also did not attempt to quantify the effect of metal NP on molecules' (or QDs) internal transitions and relaxation times which may play important role at very close distances.\cite{tomasi-jcp03,tomasi-jcp04}

\section{Conclusions}
\label{sec:conc}

In summary, were performed microscopic calculations of radiative and nonradiative decay rates of a fluorophore molecule situated near a small Au nanoparticle in dielectric medium  within quantum three-region model that incorporates non-local and quantum-size effects and accounts for different density profiles of d-band and sp-band electrons. We found that, close to metal surface, the energy transfer rate from molecular dipole to nanoparticle is significantly (by order of magnitude) higher than that predicted by semiclassical electromagnetic models due to direct excitation of electron-hole pairs by dipole's Coulomb potential. However, the non-local and finite-size effects lead to a considerably slower dependence of energy transfer rate on molecule-NP distance than $d^{-4}$ behavior predicted by RPA calculations.

This work was supported in part by the NSF under Grants No. DMR-0906945 and No. HRD-0833178, and by the EPSCOR program.

\end{document}